\documentclass[pre,aps,amsmath,amssymb,amsfonts,floatfix,superscriptaddress,showpacs,twocolumn]{revtex4-1}
\usepackage{graphicx}
\usepackage{color}
\usepackage{bbm}
\usepackage[utf8]{inputenc}
\usepackage{dcolumn}
\usepackage{hyperref}
\usepackage[caption=false]{subfig}
\usepackage[english]{babel}
\usepackage{comment}
\usepackage{indentfirst}
\hypersetup{colorlinks=true,breaklinks,linkcolor=black,urlcolor=blue,citecolor=blue}

\usepackage{tikz}
\usetikzlibrary{calc}
\usetikzlibrary{decorations.pathmorphing}
\usetikzlibrary{decorations.pathreplacing}

\renewcommand{\i}{{\ensuremath{i}}}

\newcommand{\iom}{{\ensuremath{\i\omega}}}
\newcommand{\inu}{{\ensuremath{\i\nu}}}
\newcommand{\kv}{\ensuremath{\mathbf{k}}}
\newcommand{\KV}{\ensuremath{\mathbf{q}}}

\newcommand{\qv}{\ensuremath{\mathbf{q}}}

\newcommand{\abs}[1]{\ensuremath{\lvert#1\rvert}}

\newcommand{\av}[1]{\ensuremath{\left\langle #1 \right\rangle}}

\newcommand{\up}{\ensuremath{\uparrow}}
\newcommand{\dn}{\ensuremath{\downarrow}}

\usepackage{tikz}
\usetikzlibrary{calc}
\usetikzlibrary{decorations.pathmorphing}
\usetikzlibrary{decorations.pathreplacing}
\usetikzlibrary{decorations.markings}

\tikzset{->-/.style={decoration={
  markings,
  mark=at position #1 with {\arrow{>}}},postaction={decorate}}}
\tikzset{-<-/.style={decoration={
  markings,
  mark=at position #1 with {\arrow{<}}},postaction={decorate}}}

\def \Im {\mathop {\rm Im}}
\def \Re {\mathop {\rm Re}}

\begin{document}

\title{Plasmons in Strongly Correlated Systems: Spectral Weight Transfer\\ and Renormalized Dispersion}

\author{E. G. C. P. van Loon}
\affiliation{Radboud University Nijmegen, Institute for Molecules and Materials, NL-6525 AJ Nijmegen, The Netherlands}

\author{H. Hafermann}
\affiliation{Institut de Physique Th\'eorique (IPhT), CEA, CNRS, 91191 Gif-sur-Yvette, France}

\author{A. I. Lichtenstein}
\affiliation{I. Institut f\"ur Theoretische Physik, Universit\"at Hamburg, Jungiusstra\ss e 9, D-20355 Hamburg, Germany}
\affiliation{Department of Theoretical Physics and Applied Mathematics, Ural Federal University, Mira Street 19, 620002, Ekaterinburg, Russia}

\author{A. N. Rubtsov}
\affiliation{Department of Physics, Moscow State University, 119991 Moscow, Russia}
\affiliation{Russian Quantum Center, Novaya Street 100, Skolkovo, Moscow region, 143025, Russia}

\author{M. I. Katsnelson}
\affiliation{Radboud University Nijmegen, Institute for Molecules and Materials, NL-6525 AJ Nijmegen, The Netherlands}
\affiliation{Department of Theoretical Physics and Applied Mathematics, Ural Federal University, Mira Street 19, 620002, Ekaterinburg, Russia}

\begin{abstract}
We study the charge-density dynamics within the two-dimensional extended Hubbard model in the presence of long-range Coulomb interaction across the metal-insulator transition point. 
To take into account strong correlations we start from self-consistent extended dynamical mean-field theory and include nonlocal dynamical vertex corrections through a ladder approximation to the polarization operator.
This is necessary to fulfill charge conservation and to describe plasmons in the correlated state.
The calculated plasmon spectra are qualitatively different from those in the random-phase approximation: they exhibit a spectral density transfer and a renormalized dispersion with enhanced deviation from the canonical $\sqrt{q}$ behavior.
Both features are reminiscent of interaction induced changes found in single-electron spectra of strongly correlated systems.
\end{abstract}


\maketitle

\clearpage

Substantial progress has been made in recent years, both theoretically and experimentally, in the study of the electronic structure of strongly correlated systems~\cite{Imada98}.
Such systems combine atomiclike features, such as the formation of local moments and multiplets, with the Bloch character of states of itinerant electrons. 
This requires a development of special tools for their treatment~\cite{Lichtenstein98}. 
The introduction of the dynamical mean-field theory (DMFT)~\cite{Metzner89,Georges96} was a crucial step forward since it provides a natural interpolation between atomic and band limits.
In DMFT, all local correlation effects are taken into account via a frequency-dependent local self-energy. Key phenomena described by DMFT are the spectral weight transfer, i.e., the formation of Hubbard bands~\cite{Hubbard63,Hubbard64}, the band renormalization and the associated mass enhancement, as well as the Mott transition~\cite{Georges96,Imada98}.
Since its introduction, DMFT has been extended in various ways to describe the impact of the nonlocal (intersite) correlations on the electronic structure. Examples include the quantum cluster approaches~\cite{Maier05} and the diagrammatic extensions of DMFT, such as the D$\Gamma$A~\cite{Toschi07}, dual fermion~\cite{Rubtsov08} and one-particle irreducible approach~\cite{Rohringer13}. 

Our knowledge of the {\it collective} excitations in strongly correlated systems, on the other hand, has not progressed as much. It is known that plasmons are described by the Lindhardt dielectric function within the random phase approximation (RPA)~\cite{Pines66,Platzman73,Vonsovsky89}. 
The RPA, however, is not applicable to correlated systems since plasmons are considered as a superposition of electron-hole pairs of {\it bare} electrons with unrenormalized energy spectrum. The simplest approach to correlated electrons is the $GW$ approximation. $GW$ self-energies are  obtained from a $\Psi$~functional written in terms of the bosonic and fermionic Green's functions~\cite{Almbladh99}. The self-consistent second-order variation of the functional is needed to calculate the plasmon excitation spectra and is equivalent to the solution of the Bethe-Salpeter equation~\cite{Onida02}. Because of its simple perturbative nature, it does not capture the Mott transition.
Early works on extended DMFT (EDMFT)~\cite{Si96,Parcollet99,Smith00,Chitra01,Sun03} aimed at simultaneously accounting for screening due to long-range (in particular Coulomb) interaction and Mott physics. The effect of screening is taken into account through a local retarded interaction. Including the lowest-order nonlocal diagram corrections yields the so-called EDMFT+$GW$ approach~\cite{Sun02,Biermann03,Ayral12,Ayral13}.

The charge-density dynamics is encoded in the lattice charge susceptibility $X_{E}(\qv)$, where $E$ and $\qv$ denote energy and momentum, respectively. In terms of the polarization operator $\Pi$, we can write it in the form $X_{E}(\qv) = [\Pi_{E}(\qv)^{-1} + V({\qv})]^{-1}$, where $V(\qv)$ is the Coulomb potential.
The dispersion of the collective charge excitations is determined by the poles of the charge susceptibility, i.e., by the equation
\begin{align}
\label{eqn:disp}
1 + V({\qv})\Pi_{E}(\qv)=0.
\end{align}
The standard RPA analysis is based on the fact that, in the long-wavelength limit, the polarization operator behaves as $\Pi_{E}(\qv)\sim q^{2}/E^{2}$. This property is a consequence of gauge invariance and local charge conservation (for a recent discussion, see Refs.~\cite{Rubtsov12,Hafermann14-2}).
In two dimensions (2D), the potential decays as $V(\qv)\sim 1/q$, which leads to the ``classical'' $\omega_{p}(\qv)\propto \sqrt{q}$ behavior of the plasmon dispersion with proportionality factor $\sqrt{2\pi e^2 n/m}$, with $n$ being the carrier density, $e$ and $m$ the charge and mass of the electron.

In EDMFT, both the single-particle self-energy $\Sigma$ and the polarization operator $\Pi$ are supposed to be local. While the former assumption yields a description of strong correlation physics \`a la DMFT, the latter therefore leads to unphysical behavior of the collective charge excitations and to a divergence of the excitation energy in the long-wavelength limit~\cite{Hafermann14-2}.
In Ref.~\onlinecite{Rubtsov12}, the dual boson approach has been introduced as a diagrammatic extension of EDMFT. It allows us to restore the momentum dependence of the polarization operator through a ladder summation of diagrams. This scheme includes vertex corrections beyond the EDMFT+$GW$ approximation. It is a minimal conserving approximation for correlated systems, similar to the RPA being the minimal theory for the Fermi gas. In this Letter, we employ this approach to study the charge excitations in two-dimensional correlated systems with long-range Coulomb interaction.

We proceed with the prototypical model of a strongly correlated system, the (extended) Hubbard model~\cite{Hubbard63,Hubbard64,Gutzwiller63,Kanamori63}
\begin{align}
\label{eq:H}
  H &= - t \sum_{ij} c^\dagger_{i\sigma} c^{\phantom{\dagger}}_{j\sigma}
  + \frac{1}{2} \sum_{\KV} V(\KV) \rho_\KV \rho_{-\KV},
\end{align}
on the two-dimensional square lattice with the charge susceptibility $X_{\omega}(\KV) = \av{\rho\rho}_{\omega\KV}$. In the above, $c^\dagger_{i\sigma}$ and $c^{\phantom{\dagger}}_{i\sigma}$ denote the creation and annihilation of an electron on lattice site $i$ with spin $\sigma=\up,\dn$ and
$\rho_i=c^\dagger_{i\up} c^{\phantom{\dagger}}_{i\up}+c^\dagger_{i\dn} c^{\phantom{\dagger}}_{i\dn}-1$ describes the deviation of the density at site $i$ from its average value $1$ for the half-filled case that we consider. $t$ is the hopping parameter. 

The long-range Coulomb interaction $V(\KV)$ has the form $U+V_0/\abs{\KV}$ for $\abs{\qv}>0$~\footnote{The homogeneous part $V(\KV=0)$ of the interaction vanishes because of the presence of a homogeneous background charge~\cite{Pines66,Platzman73,Vonsovsky89}.}, where $V_{0}$ is the strength of the bare or effective screened interaction. Correlated adatom systems on semiconductor surfaces~\cite{Hansmann13} and the plasmonics of graphene~\cite{Grigorenko12,Kotov12} provide beautiful examples of real phenomena which may be described within this model. The momentum dependent part corresponds to the asymptotic behavior of the Coulomb interaction in two dimensions for distances considerably larger than the interatomic distance and will result in a plasmonic branch~\cite{Pines66,Platzman73,Vonsovsky89}. For such distances, the long-range interaction is screened by the substrate and the potential strength is given by $V_0 =  2\pi e^2/\kappa$, where $\kappa$ is the dielectric constant of the substrate~\cite{Hansmann13}. At short distances, screening effects of the substrate are often negligible. 
We therefore add a variable interaction $U$ to the local part which yields an effective local interaction $U^\ast=U+\sum_{\qv}V_{0}/\abs{\qv}$.

In our calculations, we choose $4t=1$ as the energy unit and work at fixed temperature $T=0.02$ and $V_0=2$, while varying the parameter $U^{*}=U+1.1$. 
In each case, we start from a standard, self-consistent EDMFT calculation. A hybridization expansion continuous-time quantum Monte Carlo solver~\cite{Werner06,Hafermann13} with improved estimators~\cite{Hafermann14} is used to compute the imaginary-time correlation functions of the impurity model without approximation. In the final impurity solver step, we additionally compute the (reducible) impurity vertex function $\gamma_{\nu\nu'\omega}$ in the charge channel, where $\nu$($\omega$) are fermionic (bosonic) Matsubara frequencies.
The polarization operator is represented in the form~\cite{Rubtsov12}
$\Pi^{-1}_{\omega}(\qv)=[\chi_{\omega}+\chi_{\omega}\tilde{\Pi}_{\omega}(\qv)\chi_{\omega}]^{-1}-U_{\omega}$, where $\chi_{\omega}$ denotes the local charge susceptibility.
The retarded interaction $U_{\omega}$, which is the same as in EDMFT, describes the mean field screening of the local interaction. It contains the local part $U^{*}$ of the interaction and is treated on the level of the impurity model.

\begin{figure}[t]
\subfloat{a)
\includegraphics{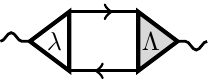}
%
%
%
%
%
%
%
}
\subfloat{b)
\includegraphics{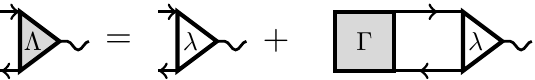}
}
\caption{\label{fig:diagrams}a) Polarization correction diagram for $\tilde{\Pi}$ and b) renormalized triangular electron-boson vertex $\Lambda$ (shaded triangle) in the dual boson approximation. The shaded square denotes the renormalized two-particle vertex $\Gamma$.}
\end{figure}

The dual bosonic self-energy $\tilde{\Pi}$ in turn is given by [cf. Fig.~\ref{fig:diagrams} a)] 
$
\tilde{\Pi}_{\omega}(\qv) = \sum_{\nu \sigma}\lambda_{\nu+\omega,-\omega}\tilde{X}^{0}_{\nu\omega}(\qv)\Lambda_{\nu\omega}(\qv)
$
(in EDMFT, it is identically zero). Here $\tilde{X}^{0}$ denotes the \emph{nonlocal part} of the bubble.
The vertex corrections in the dual boson approach enter through the renormalized triangular vertex [Fig.~\ref{fig:diagrams} b)]:
$
\Lambda_{\nu\omega}(\qv)=\lambda_{\nu\omega}+\sum_{\nu'}\Gamma_{\nu\nu'\omega}(\qv)\tilde{X}^{0}_{\nu'\omega}(\qv)\lambda_{\nu'\omega},
$
where $\Gamma_{\nu\nu'\omega}(\qv)$ denotes the lattice vertex function in the particle-hole charge channel, which is obtained through the dual Bethe-Salpeter equation~\cite{Rubtsov12}. The Bethe-Salpeter equation generates ladder diagrams, which describe the repeated particle-hole scattering processes that give rise to the long-wavelength collective excitations.

\begin{figure}[b]
\includegraphics[width=\columnwidth]{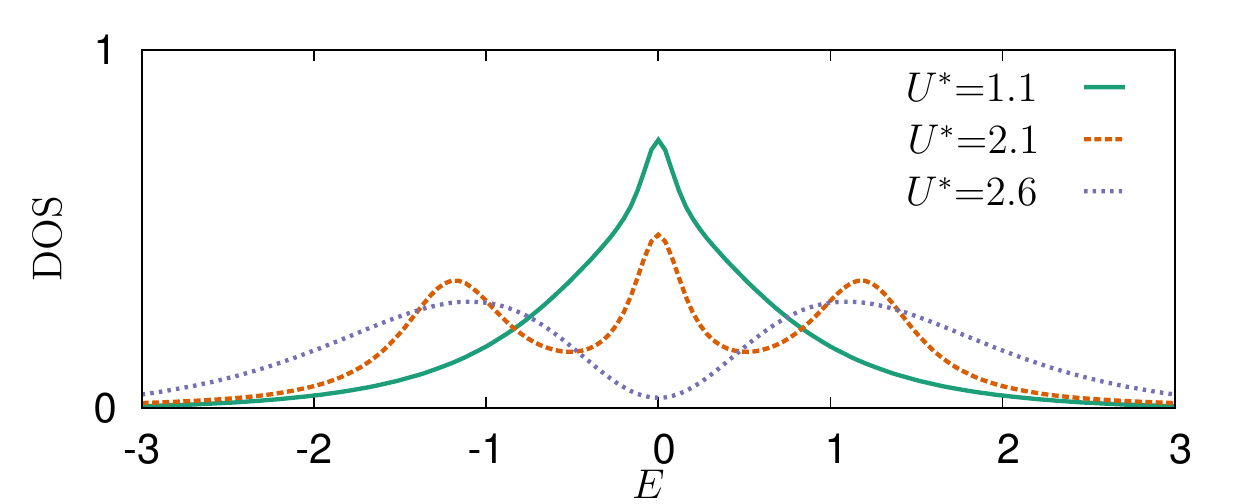}
\caption{Finite temperature local density of states of the two-dimensional Hubbard model with long-range Coulomb interaction calculated within EDMFT. The local interaction $U^{*}$ moves spectral weight from the quasiparticle peak at the Fermi energy to the Hubbard bands at $E\sim\pm U^\ast/2$. For sufficiently large $U^{*}$, the system is a Mott insulator.}
\label{fig:dos}
\end{figure}

In Fig.~\ref{fig:dos} we show the EDMFT local density of states (DOS) for the extended Hubbard model for three qualitatively different cases: For weak interaction $U^{*}$, the DOS exhibits a single quasiparticle peak at the Fermi level. As the interaction is increased, the peak is renormalized, as spectral weight is moved to incoherent excitations at a higher energy. This leads to the formation of Hubbard bands at energies $E\sim\pm U^\ast/2$.
Above a critical $U_{c}^{*}\sim 2.4$, the system undergoes a first-order Mott transition~\cite{Imada98,Georges96}. The Hubbard bands persist in the Mott phase.

\begin{figure}[!t]
\includegraphics[width=\columnwidth]{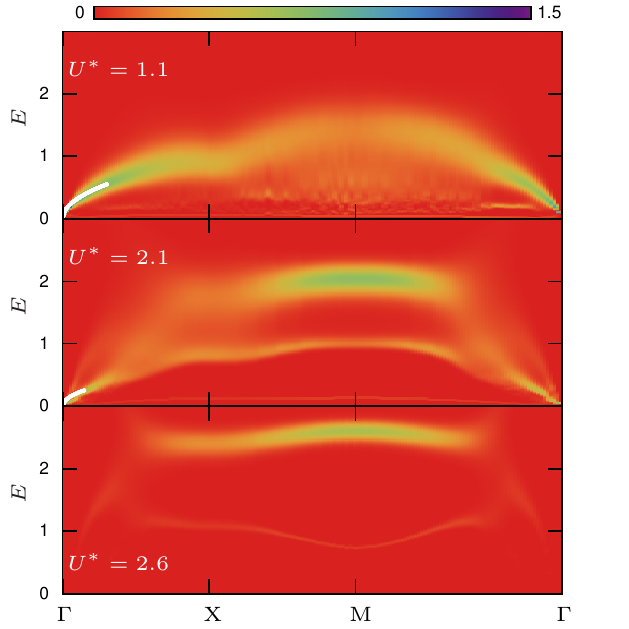}\caption{\label{fig:eels}Inverse dielectric function $-\Im \epsilon^{-1}_{E}(\KV)$ of the 2D Hubbard model with long-range Coulomb interaction for different values of the effective local interaction $U^{*}$ across the Mott transition. The spectra show a transition from itinerant to localized behavior. The interaction causes a spectral weight transfer as well as a renormalization of the long-wavelength plasmon dispersion. The dispersion relation $\omega_p(\KV)^2=\alpha V_0 q$ is shown in white.}
\end{figure}

We now turn to the discussion of the results for the collective excitations. First, we verify  in our numerical data that the polarization behaves as $q^{2}/(\iom)^{2}$ for any finite Matsubara frequency $\omega_{m}>0$ and small momenta~\cite{Hafermann14-2}. We hence expect that an RPA-type analysis according to \eqref{eqn:disp} still holds.
In Fig.~\ref{fig:eels} we plot the inverse of the dielectric function $\epsilon_{E}(\qv)=1+V(\qv)\Pi_{E}(\qv)$ as a function of real energy obtained by a stochastic analytical continuation procedure~\cite{[{Similar to the approach described in }][{. The analytical continuation by Pad\'e approximants tends to give a single peak spectrum when applied to data afflicted with statistical errors, see }]Mishchenko00,*Huang14}. Despite the appearance of artifacts, the main qualitative features discussed here are robust. This quantity can be measured via angular resolved electron energy loss spectroscopy (EELS)~\cite{Platzman73,Vonsovsky89}. We find that at $U^\ast = 1.1$, the overall behavior is reminiscent of the RPA. It can be understood within an itinerant electron picture. In the proximity of the $\Gamma$ point, the spectrum exhibits a single plasmon branch whose energy vanishes in the long-wavelength limit. We further see a rather well-defined but broadened excitation throughout the Brillouin zone with a wide continuum of particle-hole excitations below. The energy scales are determined by the hopping rather than the interaction, just as one would expect from a simple convolution of weakly renormalized Green's functions: The dominant contribution at the M-point stems from the $k$-points connected by M for which the single-particle dispersion $t_{\kv}$ is extremal, corresponding to a large density of states. The energy difference is $t_{\text{M}}-t_{\Gamma}=2$. At the X-point, $t_{\text{X}}-t_{\Gamma}=t_{\text{M}}-t_{\text{X}}=1$.  
The energy scales are indeed found to be independent of $U^{*}$ in this regime.

At $U^\ast=2.1$, the picture has changed drastically. The excitations of the particle-hole continuum are suppressed. More strikingly, the dispersion is split into two branches except for small wave vectors. The maximum energy of both branches is found at the M-point, where they are separated by a gap. The maximum of the lower branch is consistent with a value of $U^{*}/2$, while the latter is located at $E\sim U^{*}$. These features appear concomitantly with the Hubbard bands in the density of states of Fig.~\ref{fig:dos}. One can interpret the lower branch to originate mainly from particle-hole excitations for which the electron is excited from the Hubbard band to the quasiparticle-peak (or vice versa), whereas the upper branch stems from excitations between the Hubbard bands~\cite{Basov11}. Similar splitting has also been observed in EDMFT+$GW$ calculations for the extended Hubbard model with short-range interaction~\cite{Ayral12,Ayral13}.
It is further apparent that the low energy, long wavelength excitations are renormalized in the vicinity of the $\Gamma$ point. This is expected for excitations from within the quasiparticle peak. We will discuss this feature in more detail below. 
The renormalization near the $\Gamma$ point and the splitting into two branches at the M point are signs of a crossover from delocalized electrons in the metallic phase to the local physics in the insulating phase. 
At small energy scales electrons are characterized by their quasimomentum $\kv$ and the Landau picture of interacting quasiparticles is valid at long wavelengths, so there is a plasmon branch in the metal.
At larger energy scales, the whole density of states is relevant, including the Hubbard bands. Physically, the latter reflects atomiclike, localized features of strongly correlated electrons.

\begin{figure}[b]
\includegraphics[width=\columnwidth]{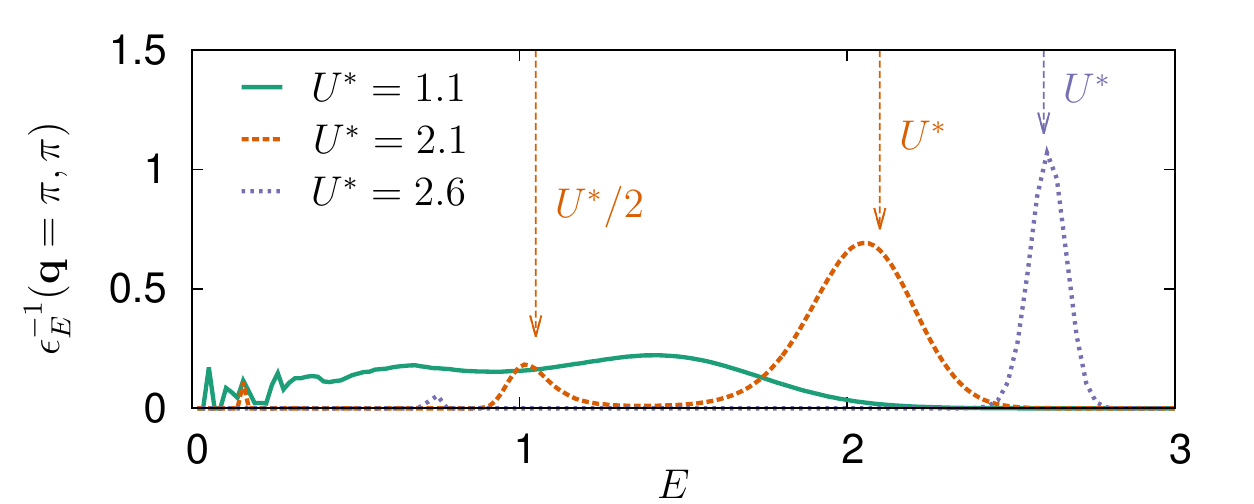}
\caption{A cross section of the EELS ($-\Im \epsilon^{-1}$) of Fig. \ref{fig:eels} at the M point, $\KV=(\pi,\pi)$. The interaction causes a transfer of spectral density.
The arrows indicate the typical energy scales $U^\ast$ and $U^\ast/2$.
}
\label{fig:spectransfer}
\end{figure}

Figure \ref{fig:eels} finally shows the inverse dielectric function in the Mott insulator at $U^\ast=2.6$. In this state, a two-particle excitation corresponds to a creation of a doublon and a holon, which costs an energy $U^\ast$. Such an excitation is expected to be highly localized. As a result, we see a weakly dispersing branch at an energy $E\sim U^{*}$. The low-energy plasmon mode has disappeared together with the quasiparticle peak.

Similar to the single-particle spectra of correlated systems, we observe a spectral weight transfer in the plasmon spectra. This is clearly illustrated in Fig.~\ref{fig:spectransfer}, where we show the inverse dielectric function for fixed momentum. The respective values of $U^{*}$ are indicated, showing that the spectral weight transfer is indeed associated with this energy scale.

The interaction also has significant impact on the plasmon dispersion. In graphene or in other systems with nonparabolic energy bands, there is a rescaling of the plasmon spectrum~\cite{Levitov13}. Evidence for a correlation-induced departure from the $\sqrt{q}$ behavior at finite $q$ has been reported in Ref.~\cite{Hirjibehedin02}. Fermi liquid theory allows a renormalization of the plasmon dispersion in systems with broken Galilean invariance, but can only provide predictions in the long wavelength limit. We now illustrate how the interaction renormalizes the dispersion for finite wave vectors.

At small $\qv$ the polarization operator is, to lowest order, proportional to $(\qv/E)^2$: $\Pi_E(\KV) = -\alpha (\qv/E)^2+\ldots$ (see Appendix for additional details). The plasmon dispersion can then be expressed using Eq.~\eqref{eqn:disp}:
\begin{align}
\omega^2_p(\KV) = \alpha V_0 q+\ldots\label{eqn:wp}.
\end{align}

We find that the effect of the interaction is to significantly lower the value of $\alpha$ and thus has two main effects:
First, the value of the plasma frequency $\omega_{0}(\qv)$ is renormalized: For the RPA with $U^{*}=1.1$ and the dual boson results with $U^{*}=1.1$ and $U^{*}=2.1$, we find $\alpha_{\text{RPA}}\approx 0.2, \alpha_{U^\ast=1.1}= 0.16 \pm 0.02$ and $\alpha_{U^\ast=2.1}\approx 0.07 \pm 0.03$, respectively.
Compared to RPA, the plasma frequency is hence renormalized for small $q$ by a factor $\sqrt{\alpha_{U^\ast=1.1}/\alpha_{\text{RPA}}}\sim 0.9$ and hence reduced by 10\% by the dynamical vertex corrections included \emph{beyond} the RPA. Comparing the cases $U^{*}=1.1$ ($U=0$) and  $U^{*}=2.1$ ($U=1$) we obtain $\sqrt{\alpha_{U^\ast=2.1}/\alpha_{U^\ast=1.1}}~\sim 0.66$, a decrease by more than 30\% induced by on-site correlations.
We find that while the plasma frequency decreases as the Mott transition is approached, the renormalization does \emph{not} scale with the quasiparticle weight $Z=(1 - d\Re \Sigma_{\omega}/d\omega)^{-1}$~\cite{Hafermann14-2}, contrary to what  one might naively expect for particle-hole excitations within the quasiparticle peak. This implies that plasmons are considerably affected by the incoherent (nonquasiparticle) properties of the system. This is consistent with the observation that the spectrum in the middle panel of Fig.~\ref{fig:eels} cannot be described in terms of a simple convolution of renormalized Green's functions.
Because of the internal summation over \emph{all} fermionic frequencies in the particle-hole bubble that appears between the scattering events in the Bethe-Salpeter equation, the quasiparticles and holes probe the entire energy spectrum. 
The collective excitations are hence influenced by the high energy scales (Hubbard bands).
Second, the interaction enhances the deviation from the $\omega_{p}(\qv)\sim\sqrt{q}$ dispersion. This can be seen in Fig.~\ref{fig:disp}, where we plot the plasmon dispersion~\eqref{eqn:wp} (solid lines)~\cite{[{See Appendix, which includes }] [{}]Vidberg77} together with the pure $\omega^2_{p} =\alpha V_0 q$ form. The interaction clearly causes the dispersion to depart from the $\sqrt{q}$ behavior at significantly smaller wave vectors.

\begin{figure}[t]
\includegraphics[width=\columnwidth]{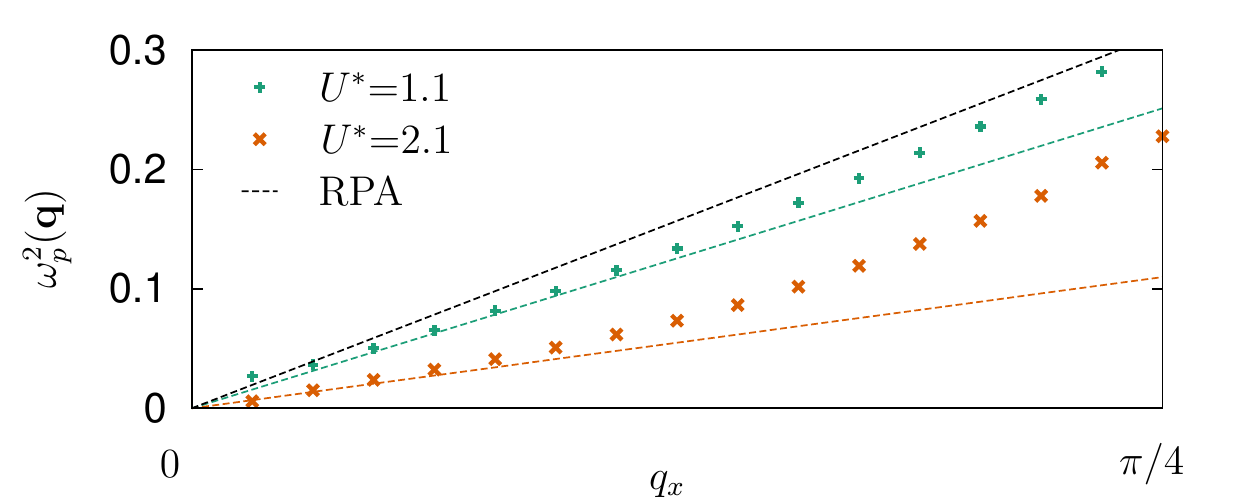}
\caption{
Renormalization of the plasmon dispersion at long wavelengths for two values of $U^{*}$. The symbols denote the dispersion obtained by analytical continuation. The dashed lines correspond to a pure square root behavior $\omega^2_{0}(\qv)=\alpha V_{0}q$. The interaction reduces the plasmon energy and causes a departure from the $\omega^2_p(\qv)\sim q$ behavior at smaller wave vectors. 
}
\label{fig:disp}
\end{figure}

We expect the spectral weight transfer and the dispersion renormalization to be general features of plasmons in strongly correlated systems. The spectral density transfer leads to qualitative differences in the EELS below and above $U_c$. For the particular case of the unfrustrated half-filled Hubbard model on the square lattice, strong antiferromagnetic correlations can also be taken into account, as they lead to a pseudogap~\cite{Rost12} and may open the gap at any finite $U$ for $T\to 0$~\cite{Schafer14}. These effects may alter our results at low temperatures. Plasmons with energies larger than the gap can, however, be expected to be unaffected by the antiferromagnetism.

In summary, we have studied the evolution of the collective charge excitations across the Mott metal-insulator transition.
By means of a ladder diagram summation, we included local as well as nonlocal vertex corrections to the polarization within the dual boson approach.
This is essential to fulfill the requirement of local charge conservation, thus providing  the $q^{2}/\omega^{2}$ behavior of the polarization operator in the long wavelength limit. It allows us to describe the collective excitations in the correlated state.
We have shown that an RPA-type analysis of the long wavelength excitations remains valid in the correlated regime. The dispersion of these long wavelength excitations is however strongly renormalized by the interaction. The renormalization is found not to scale with the quasiparticle weight, showing that plasmons are affected by incoherent properties in an essential way. Strong correlations further lead to a spectral weight transfer of the plasmonic modes, which cannot be described within the RPA. Spectral weight transfer and dispersion renormalization provoke an association of the analogous phenomena observed in single-particle spectra of strongly correlated systems.
Theoretical and experimental studies of plasmons in correlated materials, for example focusing on the mutual interplay between charge and spin degrees of freedom like in spintronics or multiferroics, are potentially relevant for applications.
It would further be very interesting to measure these effects in angular resolved electron energy loss spectroscopy~\cite{Cupolillo13} or inelastic x-ray scattering spectroscopy~\cite{Schulke01} of correlated surface systems.

The authors thank Olivier Parcollet, Thomas Ayral and Silke Biermann for valuable discussions. The work is supported by European Research Council (ERC) Advanced Grant No.~338957 FEMTO/NANO and Deutsche Forschungsgemeinschaft (DFG) Grant No.~FOR1346. A.R. acknowledges support from Dynasty and Russian Foundation for Basic Research Grant No.~14-02-01219. foundations and H.H. support from the FP7/ERC, under Grant Agreement No.~278472-MottMetals. The simulations employed a modified version of an open source implementation of the hybridization expansion quantum impurity solver~\cite{Hafermann13}, based on the ALPS libraries~\cite{ALPS2}.

\bibliography{main}

\begin{thebibliography}{47}%
\makeatletter
\providecommand \@ifxundefined [1]{%
 \@ifx{#1\undefined}
}%
\providecommand \@ifnum [1]{%
 \ifnum #1\expandafter \@firstoftwo
 \else \expandafter \@secondoftwo
 \fi
}%
\providecommand \@ifx [1]{%
 \ifx #1\expandafter \@firstoftwo
 \else \expandafter \@secondoftwo
 \fi
}%
\providecommand \natexlab [1]{#1}%
\providecommand \enquote  [1]{``#1''}%
\providecommand \bibnamefont  [1]{#1}%
\providecommand \bibfnamefont [1]{#1}%
\providecommand \citenamefont [1]{#1}%
\providecommand \href@noop [0]{\@secondoftwo}%
\providecommand \href [0]{\begingroup \@sanitize@url \@href}%
\providecommand \@href[1]{\@@startlink{#1}\@@href}%
\providecommand \@@href[1]{\endgroup#1\@@endlink}%
\providecommand \@sanitize@url [0]{\catcode `\\12\catcode `\$12\catcode
  `\&12\catcode `\#12\catcode `\^12\catcode `\_12\catcode `\%12\relax}%
\providecommand \@@startlink[1]{}%
\providecommand \@@endlink[0]{}%
\providecommand \url  [0]{\begingroup\@sanitize@url \@url }%
\providecommand \@url [1]{\endgroup\@href {#1}{\urlprefix }}%
\providecommand \urlprefix  [0]{URL }%
\providecommand \Eprint [0]{\href }%
\providecommand \doibase [0]{http://dx.doi.org/}%
\providecommand \selectlanguage [0]{\@gobble}%
\providecommand \bibinfo  [0]{\@secondoftwo}%
\providecommand \bibfield  [0]{\@secondoftwo}%
\providecommand \translation [1]{[#1]}%
\providecommand \BibitemOpen [0]{}%
\providecommand \bibitemStop [0]{}%
\providecommand \bibitemNoStop [0]{.\EOS\space}%
\providecommand \EOS [0]{\spacefactor3000\relax}%
\providecommand \BibitemShut  [1]{\csname bibitem#1\endcsname}%
\let\auto@bib@innerbib\@empty
\bibitem [{\citenamefont {Imada}\ \emph {et~al.}(1998)\citenamefont {Imada},
  \citenamefont {Fujimori},\ and\ \citenamefont {Tokura}}]{Imada98}%
  \BibitemOpen
  \bibfield  {author} {\bibinfo {author} {\bibfnamefont {M.}~\bibnamefont
  {Imada}}, \bibinfo {author} {\bibfnamefont {A.}~\bibnamefont {Fujimori}}, \
  and\ \bibinfo {author} {\bibfnamefont {Y.}~\bibnamefont {Tokura}},\
  }\href@noop {} {\bibfield  {journal} {\bibinfo  {journal} {Rev. Mod. Phys.}\
  }\textbf {\bibinfo {volume} {70}},\ \bibinfo {pages} {1039} (\bibinfo {year}
  {1998})}\BibitemShut {NoStop}%
\bibitem [{\citenamefont {Lichtenstein}\ and\ \citenamefont
  {Katsnelson}(1998)}]{Lichtenstein98}%
  \BibitemOpen
  \bibfield  {author} {\bibinfo {author} {\bibfnamefont {A.~I.}\ \bibnamefont
  {Lichtenstein}}\ and\ \bibinfo {author} {\bibfnamefont {M.~I.}\ \bibnamefont
  {Katsnelson}},\ }\href@noop {} {\bibfield  {journal} {\bibinfo  {journal}
  {Phys. Rev. B}\ }\textbf {\bibinfo {volume} {57}},\ \bibinfo {pages} {6884}
  (\bibinfo {year} {1998})}\BibitemShut {NoStop}%
\bibitem [{\citenamefont {Metzner}\ and\ \citenamefont
  {Vollhardt}(1989)}]{Metzner89}%
  \BibitemOpen
  \bibfield  {author} {\bibinfo {author} {\bibfnamefont {W.}~\bibnamefont
  {Metzner}}\ and\ \bibinfo {author} {\bibfnamefont {D.}~\bibnamefont
  {Vollhardt}},\ }\href {\doibase 10.1103/PhysRevLett.62.324} {\bibfield
  {journal} {\bibinfo  {journal} {Phys. Rev. Lett.}\ }\textbf {\bibinfo
  {volume} {62}},\ \bibinfo {pages} {324} (\bibinfo {year} {1989})}\BibitemShut
  {NoStop}%
\bibitem [{\citenamefont {Georges}\ \emph {et~al.}(1996)\citenamefont
  {Georges}, \citenamefont {Kotliar}, \citenamefont {Krauth},\ and\
  \citenamefont {Rozenberg}}]{Georges96}%
  \BibitemOpen
  \bibfield  {author} {\bibinfo {author} {\bibfnamefont {A.}~\bibnamefont
  {Georges}}, \bibinfo {author} {\bibfnamefont {G.}~\bibnamefont {Kotliar}},
  \bibinfo {author} {\bibfnamefont {W.}~\bibnamefont {Krauth}}, \ and\ \bibinfo
  {author} {\bibfnamefont {M.~J.}\ \bibnamefont {Rozenberg}},\ }\href {\doibase
  10.1103/RevModPhys.68.13} {\bibfield  {journal} {\bibinfo  {journal} {Rev.
  Mod. Phys.}\ }\textbf {\bibinfo {volume} {68}},\ \bibinfo {pages} {13}
  (\bibinfo {year} {1996})}\BibitemShut {NoStop}%
\bibitem [{\citenamefont {Hubbard}(1963)}]{Hubbard63}%
  \BibitemOpen
  \bibfield  {author} {\bibinfo {author} {\bibfnamefont {J.}~\bibnamefont
  {Hubbard}},\ }\href {\doibase 10.1098/rspa.1963.0204} {\bibfield  {journal}
  {\bibinfo  {journal} {Proc. R. Soc. A.}\ }\textbf {\bibinfo {volume} {276}},\
  \bibinfo {pages} {238} (\bibinfo {year} {1963})}\BibitemShut {NoStop}%
\bibitem [{\citenamefont {Hubbard}(1964)}]{Hubbard64}%
  \BibitemOpen
  \bibfield  {author} {\bibinfo {author} {\bibfnamefont {J.}~\bibnamefont
  {Hubbard}},\ }\href {\doibase 10.1098/rspa.1964.0190} {\bibfield  {journal}
  {\bibinfo  {journal} {Proc. R. Soc. A.}\ }\textbf {\bibinfo {volume} {281}},\
  \bibinfo {pages} {401} (\bibinfo {year} {1964})}\BibitemShut {NoStop}%
\bibitem [{\citenamefont {Maier}\ \emph {et~al.}(2005)\citenamefont {Maier},
  \citenamefont {Jarrell}, \citenamefont {Pruschke},\ and\ \citenamefont
  {Hettler}}]{Maier05}%
  \BibitemOpen
  \bibfield  {author} {\bibinfo {author} {\bibfnamefont {T.}~\bibnamefont
  {Maier}}, \bibinfo {author} {\bibfnamefont {M.}~\bibnamefont {Jarrell}},
  \bibinfo {author} {\bibfnamefont {T.}~\bibnamefont {Pruschke}}, \ and\
  \bibinfo {author} {\bibfnamefont {M.~H.}\ \bibnamefont {Hettler}},\ }\href
  {\doibase 10.1103/RevModPhys.77.1027} {\bibfield  {journal} {\bibinfo
  {journal} {Rev. Mod. Phys.}\ }\textbf {\bibinfo {volume} {77}},\ \bibinfo
  {pages} {1027} (\bibinfo {year} {2005})}\BibitemShut {NoStop}%
\bibitem [{\citenamefont {Toschi}\ \emph {et~al.}(2007)\citenamefont {Toschi},
  \citenamefont {Katanin},\ and\ \citenamefont {Held}}]{Toschi07}%
  \BibitemOpen
  \bibfield  {author} {\bibinfo {author} {\bibfnamefont {A.}~\bibnamefont
  {Toschi}}, \bibinfo {author} {\bibfnamefont {A.~A.}\ \bibnamefont {Katanin}},
  \ and\ \bibinfo {author} {\bibfnamefont {K.}~\bibnamefont {Held}},\ }\href
  {\doibase 10.1103/PhysRevB.75.045118} {\bibfield  {journal} {\bibinfo
  {journal} {Phys. Rev. B}\ }\textbf {\bibinfo {volume} {75}},\ \bibinfo
  {pages} {045118} (\bibinfo {year} {2007})}\BibitemShut {NoStop}%
\bibitem [{\citenamefont {Rubtsov}\ \emph {et~al.}(2008)\citenamefont
  {Rubtsov}, \citenamefont {Katsnelson},\ and\ \citenamefont
  {Lichtenstein}}]{Rubtsov08}%
  \BibitemOpen
  \bibfield  {author} {\bibinfo {author} {\bibfnamefont {A.~N.}\ \bibnamefont
  {Rubtsov}}, \bibinfo {author} {\bibfnamefont {M.~I.}\ \bibnamefont
  {Katsnelson}}, \ and\ \bibinfo {author} {\bibfnamefont {A.~I.}\ \bibnamefont
  {Lichtenstein}},\ }\href {\doibase 10.1103/PhysRevB.77.033101} {\bibfield
  {journal} {\bibinfo  {journal} {Phys. Rev. B}\ }\textbf {\bibinfo {volume}
  {77}},\ \bibinfo {pages} {033101} (\bibinfo {year} {2008})}\BibitemShut
  {NoStop}%
\bibitem [{\citenamefont {Rohringer}\ \emph {et~al.}(2013)\citenamefont
  {Rohringer}, \citenamefont {Toschi}, \citenamefont {Hafermann}, \citenamefont
  {Held}, \citenamefont {Anisimov},\ and\ \citenamefont
  {Katanin}}]{Rohringer13}%
  \BibitemOpen
  \bibfield  {author} {\bibinfo {author} {\bibfnamefont {G.}~\bibnamefont
  {Rohringer}}, \bibinfo {author} {\bibfnamefont {A.}~\bibnamefont {Toschi}},
  \bibinfo {author} {\bibfnamefont {H.}~\bibnamefont {Hafermann}}, \bibinfo
  {author} {\bibfnamefont {K.}~\bibnamefont {Held}}, \bibinfo {author}
  {\bibfnamefont {V.~I.}\ \bibnamefont {Anisimov}}, \ and\ \bibinfo {author}
  {\bibfnamefont {A.~A.}\ \bibnamefont {Katanin}},\ }\href {\doibase
  10.1103/PhysRevB.88.115112} {\bibfield  {journal} {\bibinfo  {journal} {Phys.
  Rev. B}\ }\textbf {\bibinfo {volume} {88}},\ \bibinfo {pages} {115112}
  (\bibinfo {year} {2013})}\BibitemShut {NoStop}%
\bibitem [{\citenamefont {Pines}\ and\ \citenamefont
  {Nozi{\`e}res}(1966)}]{Pines66}%
  \BibitemOpen
  \bibfield  {author} {\bibinfo {author} {\bibfnamefont {D.}~\bibnamefont
  {Pines}}\ and\ \bibinfo {author} {\bibfnamefont {P.}~\bibnamefont
  {Nozi{\`e}res}},\ }\href@noop {} {\emph {\bibinfo {title} {The Theory of
  Quantum Liquids: Normal Fermi liquids}}}\ (\bibinfo  {publisher} {W.A.
  Benjamin, Philadelphia},\ \bibinfo {year} {1966})\BibitemShut {NoStop}%
\bibitem [{\citenamefont {Platzman}\ and\ \citenamefont
  {Wolff}(1973)}]{Platzman73}%
  \BibitemOpen
  \bibfield  {author} {\bibinfo {author} {\bibfnamefont {P.~M.}\ \bibnamefont
  {Platzman}}\ and\ \bibinfo {author} {\bibfnamefont {P.~A.}\ \bibnamefont
  {Wolff}},\ }\href@noop {} {\emph {\bibinfo {title} {Waves and Interactions in
  Solid State Plasmas}}},\ Vol.~\bibinfo {volume} {13}\ (\bibinfo  {publisher}
  {Academic Press, New York},\ \bibinfo {year} {1973})\BibitemShut {NoStop}%
\bibitem [{\citenamefont {Vonsovsky}\ and\ \citenamefont
  {Katsnelson}(1989)}]{Vonsovsky89}%
  \BibitemOpen
  \bibfield  {author} {\bibinfo {author} {\bibfnamefont {S.~V.}\ \bibnamefont
  {Vonsovsky}}\ and\ \bibinfo {author} {\bibfnamefont {M.~I.}\ \bibnamefont
  {Katsnelson}},\ }\href@noop {} {\emph {\bibinfo {title} {{Quantum Solid State
  Physics}}}}\ (\bibinfo  {publisher} {Springer Verlag},\ \bibinfo {year}
  {1989})\BibitemShut {NoStop}%
\bibitem [{\citenamefont {Almbladh}\ \emph {et~al.}(1999)\citenamefont
  {Almbladh}, \citenamefont {von Barth},\ and\ \citenamefont {van
  Leeuwen}}]{Almbladh99}%
  \BibitemOpen
  \bibfield  {author} {\bibinfo {author} {\bibfnamefont {C.-O.}\ \bibnamefont
  {Almbladh}}, \bibinfo {author} {\bibfnamefont {U.}~\bibnamefont {von Barth}},
  \ and\ \bibinfo {author} {\bibfnamefont {R.}~\bibnamefont {van Leeuwen}},\
  }\href {\doibase 10.1142/S0217979299000436} {\bibfield  {journal} {\bibinfo
  {journal} {Int. J. Mod. Phys. B}\ }\textbf {\bibinfo {volume} {13}},\
  \bibinfo {pages} {535} (\bibinfo {year} {1999})}\BibitemShut {NoStop}%
\bibitem [{\citenamefont {Onida}\ \emph {et~al.}(2002)\citenamefont {Onida},
  \citenamefont {Reining},\ and\ \citenamefont {Rubio}}]{Onida02}%
  \BibitemOpen
  \bibfield  {author} {\bibinfo {author} {\bibfnamefont {G.}~\bibnamefont
  {Onida}}, \bibinfo {author} {\bibfnamefont {L.}~\bibnamefont {Reining}}, \
  and\ \bibinfo {author} {\bibfnamefont {A.}~\bibnamefont {Rubio}},\ }\href
  {\doibase 10.1103/RevModPhys.74.601} {\bibfield  {journal} {\bibinfo
  {journal} {Rev. Mod. Phys.}\ }\textbf {\bibinfo {volume} {74}},\ \bibinfo
  {pages} {601} (\bibinfo {year} {2002})}\BibitemShut {NoStop}%
\bibitem [{\citenamefont {Si}\ and\ \citenamefont {Smith}(1996)}]{Si96}%
  \BibitemOpen
  \bibfield  {author} {\bibinfo {author} {\bibfnamefont {Q.}~\bibnamefont
  {Si}}\ and\ \bibinfo {author} {\bibfnamefont {J.~L.}\ \bibnamefont {Smith}},\
  }\href {\doibase 10.1103/PhysRevLett.77.3391} {\bibfield  {journal} {\bibinfo
   {journal} {Phys. Rev. Lett.}\ }\textbf {\bibinfo {volume} {77}},\ \bibinfo
  {pages} {3391} (\bibinfo {year} {1996})}\BibitemShut {NoStop}%
\bibitem [{\citenamefont {Parcollet}\ and\ \citenamefont
  {Georges}(1999)}]{Parcollet99}%
  \BibitemOpen
  \bibfield  {author} {\bibinfo {author} {\bibfnamefont {O.}~\bibnamefont
  {Parcollet}}\ and\ \bibinfo {author} {\bibfnamefont {A.}~\bibnamefont
  {Georges}},\ }\href {\doibase 10.1103/PhysRevB.59.5341} {\bibfield  {journal}
  {\bibinfo  {journal} {Phys. Rev. B}\ }\textbf {\bibinfo {volume} {59}},\
  \bibinfo {pages} {5341} (\bibinfo {year} {1999})}\BibitemShut {NoStop}%
\bibitem [{\citenamefont {Smith}\ and\ \citenamefont {Si}(2000)}]{Smith00}%
  \BibitemOpen
  \bibfield  {author} {\bibinfo {author} {\bibfnamefont {J.~L.}\ \bibnamefont
  {Smith}}\ and\ \bibinfo {author} {\bibfnamefont {Q.}~\bibnamefont {Si}},\
  }\href {\doibase 10.1103/PhysRevB.61.5184} {\bibfield  {journal} {\bibinfo
  {journal} {Phys. Rev. B}\ }\textbf {\bibinfo {volume} {61}},\ \bibinfo
  {pages} {5184} (\bibinfo {year} {2000})}\BibitemShut {NoStop}%
\bibitem [{\citenamefont {Chitra}\ and\ \citenamefont
  {Kotliar}(2001)}]{Chitra01}%
  \BibitemOpen
  \bibfield  {author} {\bibinfo {author} {\bibfnamefont {R.}~\bibnamefont
  {Chitra}}\ and\ \bibinfo {author} {\bibfnamefont {G.}~\bibnamefont
  {Kotliar}},\ }\href {\doibase 10.1103/PhysRevB.63.115110} {\bibfield
  {journal} {\bibinfo  {journal} {Phys. Rev. B}\ }\textbf {\bibinfo {volume}
  {63}},\ \bibinfo {pages} {115110} (\bibinfo {year} {2001})}\BibitemShut
  {NoStop}%
\bibitem [{\citenamefont {Sun}\ and\ \citenamefont {Kotliar}(2003)}]{Sun03}%
  \BibitemOpen
  \bibfield  {author} {\bibinfo {author} {\bibfnamefont {P.}~\bibnamefont
  {Sun}}\ and\ \bibinfo {author} {\bibfnamefont {G.}~\bibnamefont {Kotliar}},\
  }\href {\doibase 10.1103/PhysRevLett.91.037209} {\bibfield  {journal}
  {\bibinfo  {journal} {Phys. Rev. Lett.}\ }\textbf {\bibinfo {volume} {91}},\
  \bibinfo {pages} {037209} (\bibinfo {year} {2003})}\BibitemShut {NoStop}%
\bibitem [{\citenamefont {Sun}\ and\ \citenamefont {Kotliar}(2002)}]{Sun02}%
  \BibitemOpen
  \bibfield  {author} {\bibinfo {author} {\bibfnamefont {P.}~\bibnamefont
  {Sun}}\ and\ \bibinfo {author} {\bibfnamefont {G.}~\bibnamefont {Kotliar}},\
  }\href {\doibase 10.1103/PhysRevB.66.085120} {\bibfield  {journal} {\bibinfo
  {journal} {Phys. Rev. B}\ }\textbf {\bibinfo {volume} {66}},\ \bibinfo
  {pages} {085120} (\bibinfo {year} {2002})}\BibitemShut {NoStop}%
\bibitem [{\citenamefont {Biermann}\ \emph {et~al.}(2003)\citenamefont
  {Biermann}, \citenamefont {Aryasetiawan},\ and\ \citenamefont
  {Georges}}]{Biermann03}%
  \BibitemOpen
  \bibfield  {author} {\bibinfo {author} {\bibfnamefont {S.}~\bibnamefont
  {Biermann}}, \bibinfo {author} {\bibfnamefont {F.}~\bibnamefont
  {Aryasetiawan}}, \ and\ \bibinfo {author} {\bibfnamefont {A.}~\bibnamefont
  {Georges}},\ }\href {\doibase 10.1103/PhysRevLett.90.086402} {\bibfield
  {journal} {\bibinfo  {journal} {Phys. Rev. Lett.}\ }\textbf {\bibinfo
  {volume} {90}},\ \bibinfo {pages} {086402} (\bibinfo {year}
  {2003})}\BibitemShut {NoStop}%
\bibitem [{\citenamefont {Ayral}\ \emph {et~al.}(2012)\citenamefont {Ayral},
  \citenamefont {Werner},\ and\ \citenamefont {Biermann}}]{Ayral12}%
  \BibitemOpen
  \bibfield  {author} {\bibinfo {author} {\bibfnamefont {T.}~\bibnamefont
  {Ayral}}, \bibinfo {author} {\bibfnamefont {P.}~\bibnamefont {Werner}}, \
  and\ \bibinfo {author} {\bibfnamefont {S.}~\bibnamefont {Biermann}},\ }\href
  {\doibase 10.1103/PhysRevLett.109.226401} {\bibfield  {journal} {\bibinfo
  {journal} {Phys. Rev. Lett.}\ }\textbf {\bibinfo {volume} {109}},\ \bibinfo
  {pages} {226401} (\bibinfo {year} {2012})}\BibitemShut {NoStop}%
\bibitem [{\citenamefont {Ayral}\ \emph {et~al.}(2013)\citenamefont {Ayral},
  \citenamefont {Biermann},\ and\ \citenamefont {Werner}}]{Ayral13}%
  \BibitemOpen
  \bibfield  {author} {\bibinfo {author} {\bibfnamefont {T.}~\bibnamefont
  {Ayral}}, \bibinfo {author} {\bibfnamefont {S.}~\bibnamefont {Biermann}}, \
  and\ \bibinfo {author} {\bibfnamefont {P.}~\bibnamefont {Werner}},\ }\href
  {\doibase 10.1103/PhysRevB.87.125149} {\bibfield  {journal} {\bibinfo
  {journal} {Phys. Rev. B}\ }\textbf {\bibinfo {volume} {87}},\ \bibinfo
  {pages} {125149} (\bibinfo {year} {2013})}\BibitemShut {NoStop}%
\bibitem [{\citenamefont {Rubtsov}\ \emph {et~al.}(2012)\citenamefont
  {Rubtsov}, \citenamefont {Katsnelson},\ and\ \citenamefont
  {Lichtenstein}}]{Rubtsov12}%
  \BibitemOpen
  \bibfield  {author} {\bibinfo {author} {\bibfnamefont {A.~N.}\ \bibnamefont
  {Rubtsov}}, \bibinfo {author} {\bibfnamefont {M.~I.}\ \bibnamefont
  {Katsnelson}}, \ and\ \bibinfo {author} {\bibfnamefont {A.~I.}\ \bibnamefont
  {Lichtenstein}},\ }\href {\doibase 10.1016/j.aop.2012.01.002} {\bibfield
  {journal} {\bibinfo  {journal} {Ann. Phys.}\ }\textbf {\bibinfo {volume}
  {327}},\ \bibinfo {pages} {1320} (\bibinfo {year} {2012})}\BibitemShut
  {NoStop}%
\bibitem [{\citenamefont {Hafermann}\ \emph {et~al.}(2014)\citenamefont
  {Hafermann}, \citenamefont {van Loon}, \citenamefont {Katsnelson},
  \citenamefont {Lichtenstein},\ and\ \citenamefont
  {Parcollet}}]{Hafermann14-2}%
  \BibitemOpen
  \bibfield  {author} {\bibinfo {author} {\bibfnamefont {H.}~\bibnamefont
  {Hafermann}}, \bibinfo {author} {\bibfnamefont {E.~G. C.~P.}\ \bibnamefont
  {van Loon}}, \bibinfo {author} {\bibfnamefont {M.~I.}\ \bibnamefont
  {Katsnelson}}, \bibinfo {author} {\bibfnamefont {A.~I.}\ \bibnamefont
  {Lichtenstein}}, \ and\ \bibinfo {author} {\bibfnamefont {O.}~\bibnamefont
  {Parcollet}},\ }\href {\doibase 10.1103/PhysRevB.90.235105} {\bibfield
  {journal} {\bibinfo  {journal} {Phys. Rev. B}\ }\textbf {\bibinfo {volume}
  {90}},\ \bibinfo {pages} {235105} (\bibinfo {year} {2014})}\BibitemShut
  {NoStop}%
\bibitem [{\citenamefont {Gutzwiller}(1963)}]{Gutzwiller63}%
  \BibitemOpen
  \bibfield  {author} {\bibinfo {author} {\bibfnamefont {M.~C.}\ \bibnamefont
  {Gutzwiller}},\ }\href {\doibase 10.1103/PhysRevLett.10.159} {\bibfield
  {journal} {\bibinfo  {journal} {Phys. Rev. Lett.}\ }\textbf {\bibinfo
  {volume} {10}},\ \bibinfo {pages} {159} (\bibinfo {year} {1963})}\BibitemShut
  {NoStop}%
\bibitem [{\citenamefont {Kanamori}(1963)}]{Kanamori63}%
  \BibitemOpen
  \bibfield  {author} {\bibinfo {author} {\bibfnamefont {J.}~\bibnamefont
  {Kanamori}},\ }\href@noop {} {\bibfield  {journal} {\bibinfo  {journal}
  {Prog. Theor. Phys.}\ }\textbf {\bibinfo {volume} {30}},\ \bibinfo {pages}
  {275} (\bibinfo {year} {1963})}\BibitemShut {NoStop}%
\bibitem [{Note1()}]{Note1}%
  \BibitemOpen
  \bibinfo {note} {The homogeneous part $V(\protect \ensuremath {\protect
  \mathbf {q}}=0)$ of the interaction vanishes because of the presence of a
  homogeneous background charge~\cite
  {Pines66,Platzman73,Vonsovsky89}.}\BibitemShut {Stop}%
\bibitem [{\citenamefont {Hansmann}\ \emph {et~al.}(2013)\citenamefont
  {Hansmann}, \citenamefont {Ayral}, \citenamefont {Vaugier}, \citenamefont
  {Werner},\ and\ \citenamefont {Biermann}}]{Hansmann13}%
  \BibitemOpen
  \bibfield  {author} {\bibinfo {author} {\bibfnamefont {P.}~\bibnamefont
  {Hansmann}}, \bibinfo {author} {\bibfnamefont {T.}~\bibnamefont {Ayral}},
  \bibinfo {author} {\bibfnamefont {L.}~\bibnamefont {Vaugier}}, \bibinfo
  {author} {\bibfnamefont {P.}~\bibnamefont {Werner}}, \ and\ \bibinfo {author}
  {\bibfnamefont {S.}~\bibnamefont {Biermann}},\ }\href {\doibase
  10.1103/PhysRevLett.110.166401} {\bibfield  {journal} {\bibinfo  {journal}
  {Phys. Rev. Lett.}\ }\textbf {\bibinfo {volume} {110}},\ \bibinfo {pages}
  {166401} (\bibinfo {year} {2013})}\BibitemShut {NoStop}%
\bibitem [{\citenamefont {Grigorenko}\ \emph {et~al.}(2012)\citenamefont
  {Grigorenko}, \citenamefont {Polini},\ and\ \citenamefont
  {Novoselov}}]{Grigorenko12}%
  \BibitemOpen
  \bibfield  {author} {\bibinfo {author} {\bibfnamefont {A.~N.}\ \bibnamefont
  {Grigorenko}}, \bibinfo {author} {\bibfnamefont {M.}~\bibnamefont {Polini}},
  \ and\ \bibinfo {author} {\bibfnamefont {K.~S.}\ \bibnamefont {Novoselov}},\
  }\href@noop {} {\bibfield  {journal} {\bibinfo  {journal} {Nat. Photon.}\
  }\textbf {\bibinfo {volume} {6}},\ \bibinfo {pages} {749} (\bibinfo {year}
  {2012})}\BibitemShut {NoStop}%
\bibitem [{\citenamefont {Kotov}\ \emph {et~al.}(2012)\citenamefont {Kotov},
  \citenamefont {Uchoa}, \citenamefont {Pereira}, \citenamefont {Guinea},\ and\
  \citenamefont {Castro~Neto}}]{Kotov12}%
  \BibitemOpen
  \bibfield  {author} {\bibinfo {author} {\bibfnamefont {V.~N.}\ \bibnamefont
  {Kotov}}, \bibinfo {author} {\bibfnamefont {B.}~\bibnamefont {Uchoa}},
  \bibinfo {author} {\bibfnamefont {V.~M.}\ \bibnamefont {Pereira}}, \bibinfo
  {author} {\bibfnamefont {F.}~\bibnamefont {Guinea}}, \ and\ \bibinfo {author}
  {\bibfnamefont {A.~H.}\ \bibnamefont {Castro~Neto}},\ }\href {\doibase
  10.1103/RevModPhys.84.1067} {\bibfield  {journal} {\bibinfo  {journal} {Rev.
  Mod. Phys.}\ }\textbf {\bibinfo {volume} {84}},\ \bibinfo {pages} {1067}
  (\bibinfo {year} {2012})}\BibitemShut {NoStop}%
\bibitem [{\citenamefont {Werner}\ \emph {et~al.}(2006)\citenamefont {Werner},
  \citenamefont {Comanac}, \citenamefont {de' Medici}, \citenamefont {Troyer},\
  and\ \citenamefont {Millis}}]{Werner06}%
  \BibitemOpen
  \bibfield  {author} {\bibinfo {author} {\bibfnamefont {P.}~\bibnamefont
  {Werner}}, \bibinfo {author} {\bibfnamefont {A.}~\bibnamefont {Comanac}},
  \bibinfo {author} {\bibfnamefont {L.}~\bibnamefont {de' Medici}}, \bibinfo
  {author} {\bibfnamefont {M.}~\bibnamefont {Troyer}}, \ and\ \bibinfo {author}
  {\bibfnamefont {A.~J.}\ \bibnamefont {Millis}},\ }\href {\doibase
  10.1103/PhysRevLett.97.076405} {\bibfield  {journal} {\bibinfo  {journal}
  {Phys. Rev. Lett.}\ }\textbf {\bibinfo {volume} {97}},\ \bibinfo {pages}
  {076405} (\bibinfo {year} {2006})}\BibitemShut {NoStop}%
\bibitem [{\citenamefont {Hafermann}\ \emph {et~al.}(2013)\citenamefont
  {Hafermann}, \citenamefont {Werner},\ and\ \citenamefont
  {Gull}}]{Hafermann13}%
  \BibitemOpen
  \bibfield  {author} {\bibinfo {author} {\bibfnamefont {H.}~\bibnamefont
  {Hafermann}}, \bibinfo {author} {\bibfnamefont {P.}~\bibnamefont {Werner}}, \
  and\ \bibinfo {author} {\bibfnamefont {E.}~\bibnamefont {Gull}},\ }\href
  {\doibase http://dx.doi.org/10.1016/j.cpc.2012.12.013} {\bibfield  {journal}
  {\bibinfo  {journal} {Comput. Phys. Commun.}\ }\textbf {\bibinfo {volume}
  {184}},\ \bibinfo {pages} {1280 } (\bibinfo {year} {2013})}\BibitemShut
  {NoStop}%
\bibitem [{\citenamefont {Hafermann}(2014)}]{Hafermann14}%
  \BibitemOpen
  \bibfield  {author} {\bibinfo {author} {\bibfnamefont {H.}~\bibnamefont
  {Hafermann}},\ }\href {\doibase 10.1103/PhysRevB.89.235128} {\bibfield
  {journal} {\bibinfo  {journal} {Phys. Rev. B}\ }\textbf {\bibinfo {volume}
  {89}},\ \bibinfo {pages} {235128} (\bibinfo {year} {2014})}\BibitemShut
  {NoStop}%
\bibitem [{\citenamefont {Mishchenko}\ \emph {et~al.}(2000)\citenamefont
  {Mishchenko}, \citenamefont {Prokof'ev}, \citenamefont {Sakamoto},\ and\
  \citenamefont {Svistunov}}]{Mishchenko00}%
  \BibitemOpen
  \bibfield  {author} {\bibinfo {author} {\bibfnamefont {A.~S.}\ \bibnamefont
  {Mishchenko}}, \bibinfo {author} {\bibfnamefont {N.~V.}\ \bibnamefont
  {Prokof'ev}}, \bibinfo {author} {\bibfnamefont {A.}~\bibnamefont {Sakamoto}},
  \ and\ \bibinfo {author} {\bibfnamefont {B.~V.}\ \bibnamefont {Svistunov}},\
  }\href {\doibase 10.1103/PhysRevB.62.6317} {\bibfield  {journal} {\bibinfo
  {journal} {Phys. Rev. B}\ }\textbf {\bibinfo {volume} {62}},\ \bibinfo
  {pages} {6317} (\bibinfo {year} {2000})}\BibitemShut {NoStop}%
\bibitem [{\citenamefont {Huang}\ \emph {et~al.}(2014)\citenamefont {Huang},
  \citenamefont {Ayral}, \citenamefont {Biermann},\ and\ \citenamefont
  {Werner}}]{Huang14}%
  \BibitemOpen
  \bibfield  {author} {\bibinfo {author} {\bibfnamefont {L.}~\bibnamefont
  {Huang}}, \bibinfo {author} {\bibfnamefont {T.}~\bibnamefont {Ayral}},
  \bibinfo {author} {\bibfnamefont {S.}~\bibnamefont {Biermann}}, \ and\
  \bibinfo {author} {\bibfnamefont {P.}~\bibnamefont {Werner}},\ }\href
  {\doibase 10.1103/PhysRevB.90.195114} {\bibfield  {journal} {\bibinfo
  {journal} {Phys. Rev. B}\ }\textbf {\bibinfo {volume} {90}},\ \bibinfo
  {pages} {195114} (\bibinfo {year} {2014})}\BibitemShut {NoStop}%
\bibitem [{\citenamefont {Basov}\ \emph {et~al.}(2011)\citenamefont {Basov},
  \citenamefont {Averitt}, \citenamefont {van~der Marel}, \citenamefont
  {Dressel},\ and\ \citenamefont {Haule}}]{Basov11}%
  \BibitemOpen
  \bibfield  {author} {\bibinfo {author} {\bibfnamefont {D.~N.}\ \bibnamefont
  {Basov}}, \bibinfo {author} {\bibfnamefont {R.~D.}\ \bibnamefont {Averitt}},
  \bibinfo {author} {\bibfnamefont {D.}~\bibnamefont {van~der Marel}}, \bibinfo
  {author} {\bibfnamefont {M.}~\bibnamefont {Dressel}}, \ and\ \bibinfo
  {author} {\bibfnamefont {K.}~\bibnamefont {Haule}},\ }\href {\doibase
  10.1103/RevModPhys.83.471} {\bibfield  {journal} {\bibinfo  {journal} {Rev.
  Mod. Phys.}\ }\textbf {\bibinfo {volume} {83}},\ \bibinfo {pages} {471}
  (\bibinfo {year} {2011})}\BibitemShut {NoStop}%
\bibitem [{\citenamefont {Levitov}\ \emph {et~al.}(2013)\citenamefont
  {Levitov}, \citenamefont {Shtyk},\ and\ \citenamefont
  {Feigelman}}]{Levitov13}%
  \BibitemOpen
  \bibfield  {author} {\bibinfo {author} {\bibfnamefont {L.~S.}\ \bibnamefont
  {Levitov}}, \bibinfo {author} {\bibfnamefont {A.~V.}\ \bibnamefont {Shtyk}},
  \ and\ \bibinfo {author} {\bibfnamefont {M.~V.}\ \bibnamefont {Feigelman}},\
  }\href {\doibase 10.1103/PhysRevB.88.235403} {\bibfield  {journal} {\bibinfo
  {journal} {Phys. Rev. B}\ }\textbf {\bibinfo {volume} {88}},\ \bibinfo
  {pages} {235403} (\bibinfo {year} {2013})}\BibitemShut {NoStop}%
\bibitem [{\citenamefont {Hirjibehedin}\ \emph {et~al.}(2002)\citenamefont
  {Hirjibehedin}, \citenamefont {Pinczuk}, \citenamefont {Dennis},
  \citenamefont {Pfeiffer},\ and\ \citenamefont {West}}]{Hirjibehedin02}%
  \BibitemOpen
  \bibfield  {author} {\bibinfo {author} {\bibfnamefont {C.~F.}\ \bibnamefont
  {Hirjibehedin}}, \bibinfo {author} {\bibfnamefont {A.}~\bibnamefont
  {Pinczuk}}, \bibinfo {author} {\bibfnamefont {B.~S.}\ \bibnamefont {Dennis}},
  \bibinfo {author} {\bibfnamefont {L.~N.}\ \bibnamefont {Pfeiffer}}, \ and\
  \bibinfo {author} {\bibfnamefont {K.~W.}\ \bibnamefont {West}},\ }\href
  {\doibase 10.1103/PhysRevB.65.161309} {\bibfield  {journal} {\bibinfo
  {journal} {Phys. Rev. B}\ }\textbf {\bibinfo {volume} {65}},\ \bibinfo
  {pages} {161309} (\bibinfo {year} {2002})}\BibitemShut {NoStop}%
\bibitem [{\citenamefont {Vidberg}\ and\ \citenamefont
  {Serene}(1977)}]{Vidberg77}%
  \BibitemOpen
  \bibfield  {author} {\bibinfo {author} {\bibfnamefont {H.~J.}\ \bibnamefont
  {Vidberg}}\ and\ \bibinfo {author} {\bibfnamefont {J.~W.}\ \bibnamefont
  {Serene}},\ }\href {\doibase 10.1007/BF00655090} {\bibfield  {journal}
  {\bibinfo  {journal} {J. Low Temp. Phys.}\ }\textbf {\bibinfo {volume}
  {29}},\ \bibinfo {pages} {179} (\bibinfo {year} {1977})}\BibitemShut
  {NoStop}%
\bibitem [{\citenamefont {Rost}\ \emph {et~al.}(2012)\citenamefont {Rost},
  \citenamefont {Gorelik}, \citenamefont {Assaad},\ and\ \citenamefont
  {Bl\"umer}}]{Rost12}%
  \BibitemOpen
  \bibfield  {author} {\bibinfo {author} {\bibfnamefont {D.}~\bibnamefont
  {Rost}}, \bibinfo {author} {\bibfnamefont {E.~V.}\ \bibnamefont {Gorelik}},
  \bibinfo {author} {\bibfnamefont {F.}~\bibnamefont {Assaad}}, \ and\ \bibinfo
  {author} {\bibfnamefont {N.}~\bibnamefont {Bl\"umer}},\ }\href {\doibase
  10.1103/PhysRevB.86.155109} {\bibfield  {journal} {\bibinfo  {journal} {Phys.
  Rev. B}\ }\textbf {\bibinfo {volume} {86}},\ \bibinfo {pages} {155109}
  (\bibinfo {year} {2012})}\BibitemShut {NoStop}%
\bibitem [{\citenamefont {{Sch{\"a}fer}}\ \emph {et~al.}(2014)\citenamefont
  {{Sch{\"a}fer}}, \citenamefont {{Geles}}, \citenamefont {{Rost}},
  \citenamefont {{Rohringer}}, \citenamefont {{Arrigoni}}, \citenamefont
  {{Held}}, \citenamefont {{Bl{\"u}mer}}, \citenamefont {{Aichhorn}},\ and\
  \citenamefont {{Toschi}}}]{Schafer14}%
  \BibitemOpen
  \bibfield  {author} {\bibinfo {author} {\bibfnamefont {T.}~\bibnamefont
  {{Sch{\"a}fer}}}, \bibinfo {author} {\bibfnamefont {F.}~\bibnamefont
  {{Geles}}}, \bibinfo {author} {\bibfnamefont {D.}~\bibnamefont {{Rost}}},
  \bibinfo {author} {\bibfnamefont {G.}~\bibnamefont {{Rohringer}}}, \bibinfo
  {author} {\bibfnamefont {E.}~\bibnamefont {{Arrigoni}}}, \bibinfo {author}
  {\bibfnamefont {K.}~\bibnamefont {{Held}}}, \bibinfo {author} {\bibfnamefont
  {N.}~\bibnamefont {{Bl{\"u}mer}}}, \bibinfo {author} {\bibfnamefont
  {M.}~\bibnamefont {{Aichhorn}}}, \ and\ \bibinfo {author} {\bibfnamefont
  {A.}~\bibnamefont {{Toschi}}},\ }\href@noop {} {\  (\bibinfo {year}
  {2014})},\ \Eprint {http://arxiv.org/abs/1405.7250} {arXiv:1405.7250
  [cond-mat.str-el]} \BibitemShut {NoStop}%
\bibitem [{\citenamefont {Cupolillo}\ \emph {et~al.}(2013)\citenamefont
  {Cupolillo}, \citenamefont {Ligato},\ and\ \citenamefont
  {Caputi}}]{Cupolillo13}%
  \BibitemOpen
  \bibfield  {author} {\bibinfo {author} {\bibfnamefont {A.}~\bibnamefont
  {Cupolillo}}, \bibinfo {author} {\bibfnamefont {N.}~\bibnamefont {Ligato}}, \
  and\ \bibinfo {author} {\bibfnamefont {L.}~\bibnamefont {Caputi}},\ }\href
  {\doibase http://dx.doi.org/10.1016/j.susc.2012.09.018} {\bibfield  {journal}
  {\bibinfo  {journal} {Surf. Sci.}\ }\textbf {\bibinfo {volume} {608}},\
  \bibinfo {pages} {88 } (\bibinfo {year} {2013})}\BibitemShut {NoStop}%
\bibitem [{\citenamefont {Schülke}(2001)}]{Schulke01}%
  \BibitemOpen
  \bibfield  {author} {\bibinfo {author} {\bibfnamefont {W.}~\bibnamefont
  {Schülke}},\ }\href {http://stacks.iop.org/0953-8984/13/i=34/a=307}
  {\bibfield  {journal} {\bibinfo  {journal} {J. Phys. Condens. Matter}\
  }\textbf {\bibinfo {volume} {13}},\ \bibinfo {pages} {7557} (\bibinfo {year}
  {2001})}\BibitemShut {NoStop}%
\bibitem [{\citenamefont {Bauer}\ \emph {et~al.}(2011)\citenamefont {Bauer},
  \citenamefont {Carr}, \citenamefont {Evertz}, \citenamefont {Feiguin},
  \citenamefont {Freire}, \citenamefont {Fuchs}, \citenamefont {Gamper},
  \citenamefont {Gukelberger}, \citenamefont {Gull}, \citenamefont {Guertler},
  \citenamefont {Hehn}, \citenamefont {Igarashi}, \citenamefont {Isakov},
  \citenamefont {Koop}, \citenamefont {Ma}, \citenamefont {Mates},
  \citenamefont {Matsuo}, \citenamefont {Parcollet}, \citenamefont
  {Pawłowski}, \citenamefont {Picon}, \citenamefont {Pollet}, \citenamefont
  {Santos}, \citenamefont {Scarola}, \citenamefont {Schollwöck}, \citenamefont
  {Silva}, \citenamefont {Surer}, \citenamefont {Todo}, \citenamefont {Trebst},
  \citenamefont {Troyer}, \citenamefont {Wall}, \citenamefont {Werner},\ and\
  \citenamefont {Wessel}}]{ALPS2}%
  \BibitemOpen
  \bibfield  {author} {\bibinfo {author} {\bibfnamefont {B.}~\bibnamefont
  {Bauer}}, \bibinfo {author} {\bibfnamefont {L.~D.}\ \bibnamefont {Carr}},
  \bibinfo {author} {\bibfnamefont {H.~G.}\ \bibnamefont {Evertz}}, \bibinfo
  {author} {\bibfnamefont {A.}~\bibnamefont {Feiguin}}, \bibinfo {author}
  {\bibfnamefont {J.}~\bibnamefont {Freire}}, \bibinfo {author} {\bibfnamefont
  {S.}~\bibnamefont {Fuchs}}, \bibinfo {author} {\bibfnamefont
  {L.}~\bibnamefont {Gamper}}, \bibinfo {author} {\bibfnamefont
  {J.}~\bibnamefont {Gukelberger}}, \bibinfo {author} {\bibfnamefont
  {E.}~\bibnamefont {Gull}}, \bibinfo {author} {\bibfnamefont {S.}~\bibnamefont
  {Guertler}}, \bibinfo {author} {\bibfnamefont {A.}~\bibnamefont {Hehn}},
  \bibinfo {author} {\bibfnamefont {R.}~\bibnamefont {Igarashi}}, \bibinfo
  {author} {\bibfnamefont {S.~V.}\ \bibnamefont {Isakov}}, \bibinfo {author}
  {\bibfnamefont {D.}~\bibnamefont {Koop}}, \bibinfo {author} {\bibfnamefont
  {P.~N.}\ \bibnamefont {Ma}}, \bibinfo {author} {\bibfnamefont
  {P.}~\bibnamefont {Mates}}, \bibinfo {author} {\bibfnamefont
  {H.}~\bibnamefont {Matsuo}}, \bibinfo {author} {\bibfnamefont
  {O.}~\bibnamefont {Parcollet}}, \bibinfo {author} {\bibfnamefont
  {G.}~\bibnamefont {Pawłowski}}, \bibinfo {author} {\bibfnamefont {J.~D.}\
  \bibnamefont {Picon}}, \bibinfo {author} {\bibfnamefont {L.}~\bibnamefont
  {Pollet}}, \bibinfo {author} {\bibfnamefont {E.}~\bibnamefont {Santos}},
  \bibinfo {author} {\bibfnamefont {V.~W.}\ \bibnamefont {Scarola}}, \bibinfo
  {author} {\bibfnamefont {U.}~\bibnamefont {Schollwöck}}, \bibinfo {author}
  {\bibfnamefont {C.}~\bibnamefont {Silva}}, \bibinfo {author} {\bibfnamefont
  {B.}~\bibnamefont {Surer}}, \bibinfo {author} {\bibfnamefont
  {S.}~\bibnamefont {Todo}}, \bibinfo {author} {\bibfnamefont {S.}~\bibnamefont
  {Trebst}}, \bibinfo {author} {\bibfnamefont {M.}~\bibnamefont {Troyer}},
  \bibinfo {author} {\bibfnamefont {M.~L.}\ \bibnamefont {Wall}}, \bibinfo
  {author} {\bibfnamefont {P.}~\bibnamefont {Werner}}, \ and\ \bibinfo {author}
  {\bibfnamefont {S.}~\bibnamefont {Wessel}},\ }\href@noop {} {\bibfield
  {journal} {\bibinfo  {journal} {J. Stat. Mech.}\ }\textbf {\bibinfo {volume}
  {2011}},\ \bibinfo {pages} {P05001} (\bibinfo {year} {2011})}\BibitemShut
  {NoStop}%
\bibitem [{\citenamefont {van Loon}\ \emph {et~al.}(2014)\citenamefont {van
  Loon}, \citenamefont {Lichtenstein}, \citenamefont {Katsnelson},
  \citenamefont {Parcollet},\ and\ \citenamefont {Hafermann}}]{VanLoon14}%
  \BibitemOpen
  \bibfield  {author} {\bibinfo {author} {\bibfnamefont {E.~G. C.~P.}\
  \bibnamefont {van Loon}}, \bibinfo {author} {\bibfnamefont {A.~I.}\
  \bibnamefont {Lichtenstein}}, \bibinfo {author} {\bibfnamefont {M.~I.}\
  \bibnamefont {Katsnelson}}, \bibinfo {author} {\bibfnamefont
  {O.}~\bibnamefont {Parcollet}}, \ and\ \bibinfo {author} {\bibfnamefont
  {H.}~\bibnamefont {Hafermann}},\ }\href {\doibase 10.1103/PhysRevB.90.235135}
  {\bibfield  {journal} {\bibinfo  {journal} {Phys. Rev. B}\ }\textbf {\bibinfo
  {volume} {90}},\ \bibinfo {pages} {235135} (\bibinfo {year}
  {2014})}\BibitemShut {NoStop}%
\end{thebibliography}%

\appendix

\clearpage
\onecolumngrid
\setcounter{table}{0}
\setcounter{section}{0}
\setcounter{figure}{0}
\setcounter{equation}{0}
\renewcommand{\thepage}{\Roman{page}}
\renewcommand{\thefigure}{S\arabic{figure}}
\renewcommand{\theequation}{S\arabic{equation}}

In this supplement we will

\begin{itemize}
 \item Explain the dual boson computational scheme. 
 \item Study the polarization operator in the limit $q\rightarrow 0$.
 \item Explain how to obtain the plasmon dispersion.
\end{itemize}

\section{Dual Boson method}
\label{sec:db}

The dual boson approach is based on a seperation of a many-body lattice problem into local and non-local parts. For a derivation of the approach, we refer the reader to Ref. \onlinecite{Rubtsov12}. For a recent discussion, see also Refs. \onlinecite{Hafermann14-2} and \onlinecite{VanLoon14}. Here we will briefly summarize how the polarization operator and susceptibility are calculated.

As in DMFT and EDMFT, the dual boson method is based on the self-consistent solution of an impurity problem. This impurity problem is described by a hybridization function $\Delta_\nu$ and a frequency-dependent interaction $U_\omega$:
\begin{align}
\label{simp}
S_{\text{imp}}[c^{*},c]=&-\sum_{\nu\sigma} c^{*}_{\nu\sigma}[\inu+\mu-\Delta_{\nu}]c_{\nu\sigma}+ \frac{1}{2}\sum_{\omega}U_{\omega} \rho_{\omega} \rho_{-\omega},
\end{align}
In the summations over the bosonic and fermionic Matsubara frequencies $\omega$, $\nu$ (or quasimomenta), we implicitly assume a normalization by the inverse temperature (or number of $\kv$ points).
The functions $\Delta_\nu$ and $U_\omega$ are chosen in such a way that the impurity (fermionic and bosonic) Green's functions and the local Green's functions of the lattice model are identical. The self-consistency procedure is identical to EDMFT (see~\cite{Ayral13} for a recent discussion).

The impurity problem is solved numerically exactly using a continuous-time Quantum Monte Carlo solver~\cite{Werner06} that can treat the frequency-dependent charge-charge interaction~\cite{Ayral13}. We use a modified version of the open source implementation presented in Ref.~\onlinecite{Hafermann13} with improved estimators for the impurity vertex functions~\cite{Hafermann14}. The impurity solver determines the impurity Green's function $g_{\nu}$, susceptibility $
\chi_{\omega}$, fermion-fermion vertex $\gamma$ and fermion-boson vertex $\lambda$, which all enter the dual perturbation theory. The fermionic impurity Green's function $g_{\nu_n}$ is determined on fermionic Matsubara frequencies $\nu_{n}=(2n+1)\pi/\beta$ with $n\in[0, \ldots 255]$, the susceptibility $\chi_{\omega_m}$ on bosonic Matsubara frequencies $\omega_{m}=2m\pi/\beta$ with $m\in [0, \ldots 127]$ and the vertices $\gamma_{\nu_n\nu_{n'}\omega_m}$, $\lambda_{\nu_n\omega_m}$ are determined for $m\in [0, \ldots 127]$, $n,n' \in [-64, \ldots 63]$.
We only consider the paramagnetic case, so spin labels have been omitted.

From the impurity problem, we determine the nonlocal Green's function $\tilde{G}_{\nu}(\kv)=G_{\nu}(\kv)-g_{\nu}$, where $G_{\nu}(\kv)=\left[g_{\nu}^{-1} + (\Delta_{\nu}-t_{\kv})\right]^{-1}$ is the usual (E)DMFT lattice Green's function.
Then the nonlocal part of the bubble $\tilde{X}^{0}_{\nu\omega}(\qv)=-\sum_{\kv}\tilde{G}_{\nu}(\kv)\tilde{G}_{\nu+\omega}(\kv+\qv)$ is calculated and used to invert the Bethe-Salpeter equation (BSE) to obtain the renormalized vertex $\Gamma$: $
[\Gamma^{-1}_{\omega}(\qv)]_{\nu\nu'} = [\gamma^{-1}_{\omega}]_{\nu\nu'} -\tilde{X}^{0}_{\nu\omega}(\qv)\delta_{\nu\nu'}.
$
The BSE generates repeated particle-hole scattering processes to all orders, which is necessary for a correct description of collective excitations~\cite{Hafermann14-2}.
In the summations over the bosonic and fermionic Matsubara frequencies $\omega$, $\nu$ (or quasimomenta), we implicitly assume a normalization by the inverse temperature (or number of $\kv$ points). 

The dual polarization [see Fig. 1a). of the main text] is then given by
\begin{align}
\label{supp:eq:pi}
\tilde{\Pi}_{\omega}(\qv) =&\phantom{+}\sum_{\nu\sigma}\lambda_{\nu+\omega,-\omega}\tilde{X}^{0}_{\nu\omega}(\qv)\lambda_{\nu\omega}\notag \\
&+\sum_{\nu\nu'\sigma\sigma'}\lambda_{\nu+\omega,-\omega}\tilde{X}^{0}_{\nu\omega}(\qv)\Gamma_{\nu\nu'\omega}(\qv)\tilde{X}^{0}_{\nu'\omega}(\qv)\lambda_{\nu'\omega}.
\end{align}
Here the renormalized triangular vertex [see Fig. 1b) of the main text] $
\Lambda_{\nu\omega}(\qv)=\lambda_{\nu\omega}+\sum_{\nu'}\Gamma_{\nu\nu'\omega}(\qv)\tilde{X}^{0}_{\nu'\omega}(\qv)\lambda_{\nu'\omega}
$
has been inserted.

Computational efficiency can be improved by noting that the procedure is diagonal in $\omega$ and $\qv$. The calculation is hence parallelized over $\omega$ and $\Gamma_{\omega}(\qv)$ is stored for only one $\qv$ at a time. We performed calculations on a discrete $128\times 128$ lattice with periodic boundary conditions, employing lattice symmetries and Fast Fourier Transforms to reduce the computational cost. 

Finally, the physical polarization $\Pi_{\omega}(\qv)$ is determined from the dual polarization $\tilde{\Pi}_{\omega}(\qv)$,  using the relation~\cite{Rubtsov12}
\begin{align}
\Pi^{-1}_{\omega}(\qv)=[\chi_{\omega}+\chi_{\omega}\tilde{\Pi}_{\omega}(\qv)\chi_{\omega}]^{-1}-U_{\omega}.
\end{align}
For $\qv=0$, charge conservation requires $\tilde{\Pi}_{\omega}(\qv)\chi_{\omega}=-1$, which can be used to control the accuracy of a simulation. When too few fermionic frequencies are used in \eqref{supp:eq:pi}, $\tilde{\Pi}_{\omega}(\qv)\chi_{\omega}$ deviates from $-1$. Statistical noise in the impurity quantities (especially in $\gamma_{\nu\nu'\omega}$) shows up as noise in the relation $\tilde{\Pi}_{\omega}(\qv)\chi_{\omega}=-1$.

\section{Long-wavelength behavior of the polarization operator}
\label{sec:pi}

Charge conservation requires that $\Pi_{E}(\qv)\rightarrow 0$ for $\qv\rightarrow 0$ and $E>0$~\cite{Rubtsov12,Hafermann14-2}. We confirm this in our simulations. The lattice symmetry implies that $\Pi_{E}(\qv)$ is even in $\qv$, so $\Pi_{E}(\qv) \propto q^2$ for fixed $E$ and small $\qv$.
In RPA, the polarization is given by the Lindhardt function which in the long wavelength limit and for small $q/E$ behaves as $q^{2}/E^{2}$.
In the correlated case, the polarization exhibits the same discontinuity at $\qv=\omega=0$ as the Lindhardt bubble and we assume that for small $q=\abs{\qv}$ and small $q/E$, the polarization operator has the form
$
\Pi_{E}(\KV) = -\alpha (q/E)^2.
$
with $\alpha>0$. The plasmon dispersion $\omega_p(\qv)$ is determined by
$
1 + V({\qv})\Pi_{E=\omega_p(\qv)}(\qv)=0,
$ 
or $\omega_p(\qv)^2 = \alpha V(\qv) q^2$. In two dimensions, we have $V(\qv)=U+V_{0}/q$ so that to leading order in $q$, we recover the $\sqrt{q}$ behavior of the plasmon dispersion: $\omega_p(q)\approx \sqrt{\alpha V_{0} q}$.

\begin{figure}[t]
\includegraphics[width=\columnwidth]{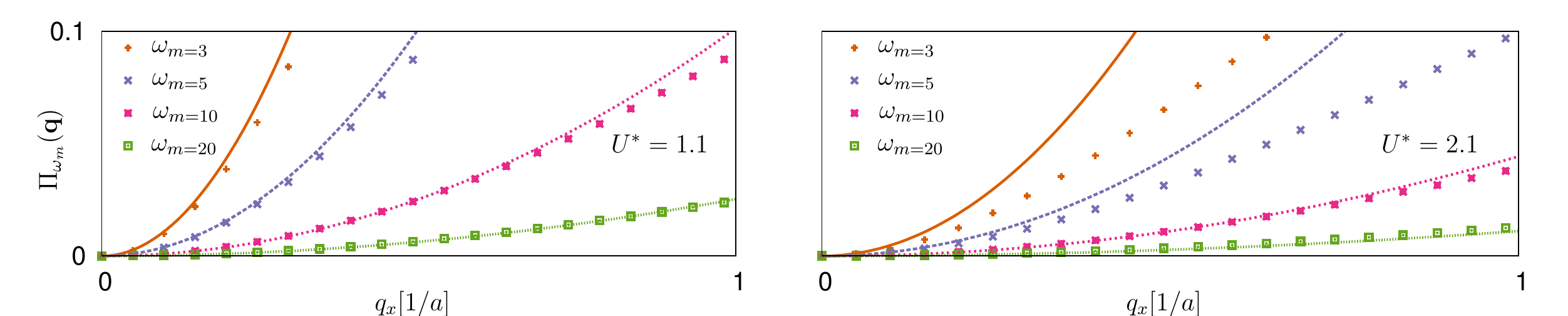}
\caption{
The long-wavelength behavior of the polarization $\Pi_{\omega}(\qv)$. The symbols denote $\Pi_{\omega}(\qv)$ on Matsubara frequencies, the lines show fits to the form $-\alpha (q/\omega)^2$, with the same $\alpha$ for fixed interaction: $\alpha_{U^\ast=1.1}=0.16$ (left) and $\alpha_{U^\ast=2.1}=0.07$ (right). The momenta lie on the first part of the path $\Gamma=(0,0)$ $\text{X}=(\pi,0)$, i.e. $q_y=0$.}
\label{fig:supp:scaling}
\end{figure}

That the polarization indeed exhibits this behavior is shown in Fig.~\ref{fig:supp:scaling}, where we show fits of the polarization operator to the form $\Pi_{\omega}(\qv)=-\alpha (q/\iom)^2$. We note that ladder diagrams at all orders contribute to the long wavelength behavior. One can further see that for small frequencies the range of momentum values for which this approximation holds is smaller, since $q/E$ should be small. Since we lack the necessary momentum resolution in this range, we fit the polarization on intermediate ($m\approx 10$) bosonic Matsubara frequencies.

The value of $\alpha$ is reduced by increasing the interaction ($\alpha_{U^\ast=1.1}\approx 0.16$ and $\alpha_{U^\ast=2.1}\approx 0.07$). Deviations from the asymptotic behavior also become important already for smaller $q$ in the case $U^\ast=2.1$. These same two effects are visible in the inverse dielectric function (see main text).

\section{Analytical continuation and plasmon dispersion}

We obtain the polarization operator, susceptibility and inverse dielectric function on Matsubara frequencies $\omega_n$. To analyze the spectrum, it is necessary to continue these quantities to real energies. However, the quantities contain two main sources of error: statistical noise from the impurity quantities and finite frequency cutoffs in the internal summation (see also \ref{sec:db}).
Since analytical continuation is inherently unstable, care is needed in the procedure.
These problems are most severe at small $q$, where the diagrammatic corrections are most relevant and in the strongly correlated regime, where where a large frequency range is needed to capture all relevant energy scales. 

We have used both Pad\'e and stochastic analytical continuation methods. We perform the analytical continuation of the susceptibility from Matsubara frequency to real energy for each $q$-point separately and then calculate the inverse dielectric function on the real axis.

The stochastic analytical continuation which was used to obtain the results of Figure 3, exhibits some artifacts. Some spectral weight appears at the edge, i.e. at the cutoff-energy (this is not visible in the plot).
Another artifact is that the spectral weight corresponding to the particle-hole continuum for $U^\ast=1.1$
comes out irregular as can be seen in Figure 4. For $U^\ast=2.1$, one can see a peak at $E=0.1$, the width of which is restricted to a single point. In the insulating $U^\ast=2.6$ spectrum, there is some spectral weight at $E\approx 0.7-0.8$, which is a factor 20 less intense than the main peak. On physical grounds we do not expect transitions with this energy in the insulator and consider this small amount of spectral weight to be an artifact.
The artifacts can be distinguished from the physical features, because the latter are robust when changing a control parameter (the real cutoff frequency in this case), while the former change their shape or position.

At small $\qv$ and in the metallic phase, the inverse dielectric function $-\Im \epsilon^{-1}_{E}(\qv)$ has a low energy plasmon mode. The dispersion of this plasmon mode can be obtained by finding, for every fixed $\qv$, the energy $E$ where $-\Im \epsilon^{-1}_{E}(\qv)$ [or, equivalently, the susceptibility $\Im X_{E}(\qv)$] has a maximum. 
In our experience, the most stable way to  do this is by using Pad\'e approximants~\cite{Vidberg77}, with a different number of Matsubara frequencies $N^{\text{Pad\'e}}_\omega = [30 \ldots 60]$, determine the dispersion for each $N^{\text{Pad\'e}}_\omega$ and finally take the average. This reduces the effect of high frequency noise. This results in the dispersion shown as points in figure 5. It is consistent with the dispersion obtained from fitting the polarization on Matsubara data (see \ref{sec:pi}) and with the maximum of the spectral weight of the spectral function in figure 3, which was obtained using stochastic analytical continuation.

\end{document}